\documentclass[aip,jcp,reprint]{revtex4-1}
\usepackage{graphicx}
\usepackage{bm,amsmath}
\usepackage{dcolumn,longtable}
\usepackage{natbib}
\usepackage[colorlinks]  
{hyperref}
\usepackage{float}
\usepackage{wrapfig}
\usepackage{ulem}

\newcommand{\fref}[1]{Fig.~\ref{#1}}

\newcommand{\sref}[1]{Sec.~\ref{#1}}
\newcommand{\Eref}[1]{Eq.~(\ref{#1})}
\newcommand{\tref}[1]{Table~\ref{#1}}

\begin{document}

\title{Sensitivity of Tunneling-Rotational Transitions in Ethylene Glycol to
Variation of Electron-to-Proton Mass Ratio}

\author{A. V. Viatkina}
\affiliation{Division of Quantum Mechanics, St.~Petersburg State University, 198904, Russia}
\author{M. G. Kozlov}
\affiliation{Petersburg Nuclear Physics Institute, Gatchina 188300, Russia}
\affiliation{St.\ Petersburg Electrotechnical University ``LETI'', Russia}
\date{
\today}

\begin{abstract}
Ethylene glycol in its ground conformation has tunneling transition with the
frequency about 7 GHz. This leads to a rather complicated tunneling-rotational
spectrum. Because tunneling and rotational energies have different dependence
on the electron-to-proton mass ratio $\mu$, this spectrum is highly sensitive
to the possible $\mu$ variation. We used simple 14 parameter effective
Hamiltonian to calculate dimensionless sensitivity coefficients $Q_\mu$ of the
tunneling-rotational transitions and found that they lie in the range from
$-17$ to $+18$. Ethylene glycol has been detected in the interstellar medium.
All this makes it one of the most sensitive probes of $\mu$ variation at the
large space and time scales.
\end{abstract}

\maketitle

\section{Introduction}\label{sec_intro}

Traditionally such values as $\alpha=\frac{e^2}{\hbar c}$ -- fine structure
constant and  $\mu=\frac{m_{e}}{m_{p}}$ -- electron to proton mass ratio are
considered unchanging over time and space. But since their exact values cannot
be calculated within the Standard model, it is natural to question their
invariability.

For the  first time this issue was addressed by Dirac in 1937 \cite{Dir37}; he
pointed out an interesting numerical coincidence (which is specific to current
age of the Universe) between two large dimensionless ratios involving
fundamental constants ($H_0$ -- Hubble constant, $c$ -- speed of light,
$\hbar$ -- Planck constant, $G$ -- gravitational constant, $\alpha , m_e,
m_p$, etc.). More precisely, Dirac noticed that the ratio of electrostatic and
gravitational attraction between a proton and an electron is the same order of
magnitude as the age of the Universe in atomic units (atomic unit of time is
$\hbar/\alpha^2c^2m_{e} \approx 2.42\times 10^{-17}$ s). He suggested that
this coincidence should persist and thus some of the involved constants have
to change over time.

Multiple other theories with slowly varying parameters appeared since then.
They connect the drift of constants with the existence of additional
dimensions in space \cite{Dam12}, or the different local density of matter
around the Universe (Chameleon theories) \cite{KW04,OP08}, or with some global
scalar field \cite{BaLi12,Uza11}. Testing these models can lead to deeper
understanding of physics.

On the contrary recent laboratory experiments
\cite{Ros08,CLN07,SBCA08,BLCT08,FJMM12,BLCT08}, astronomical observations and
geophysical evidence \cite{PNO06} have placed tight constraints on the
possible variation of $\alpha$ and $\mu$; in fact they tempt us to declare an
actual invariability of their numerical values. Current laboratory bounds (on
1$\sigma$ level) are \cite{Ros08,BLCT08}:
\begin{align}
 \label{lim_la}
 &|\dot{\alpha} / \alpha| < 4\times 10^{-17}\; \mathrm{ yr}^{-1}\,,
 \\
 \label{lim_lm}
 &|\dot{\mu} / \mu| < 3\times 10^{-15}\; \mathrm{ yr}^{-1}\,.
\end{align}
The high redshift astrophysical observations lead to the following limits
\cite{LCB12,BJHB13} ( 1$\sigma$, presuming a linear change in time):
\begin{align}
 \label{lim_aa}
 &|\dot{\alpha} / \alpha| <6\times 10^{-16}\; \mathrm{ yr}^{-1}\,,
 \\
 \label{lim_am}
 &|\dot{\mu} / \mu| < 1.5\times 10^{-17}\; \mathrm{ yr}^{-1}\,.
\end{align}
At the same time there is tentative astrophysical evidence that $\alpha$ is
changing in space (``Australian dipole'') \cite{WKM10}.

These constraints obviously put limits on theories beyond the Standard model, so that
constants should change slowly if not at all. Testing the \textquotedblleft
constancy of constants\textquotedblright\ such as $\alpha$ and $\mu$ is
examining the Einstein principle of local position invariance: \textit{``the
outcome of any local non-gravitational experiment is independent of where and
when it is being carried out.''} In order to experimentally prove or refute
invariability of constants more experiments are needed. In point of fact we
are testing the laws of physics that we are currently using, the basis of our
understanding of the Universe.


\begin{figure*}[tbh]
\includegraphics[width=\textwidth]{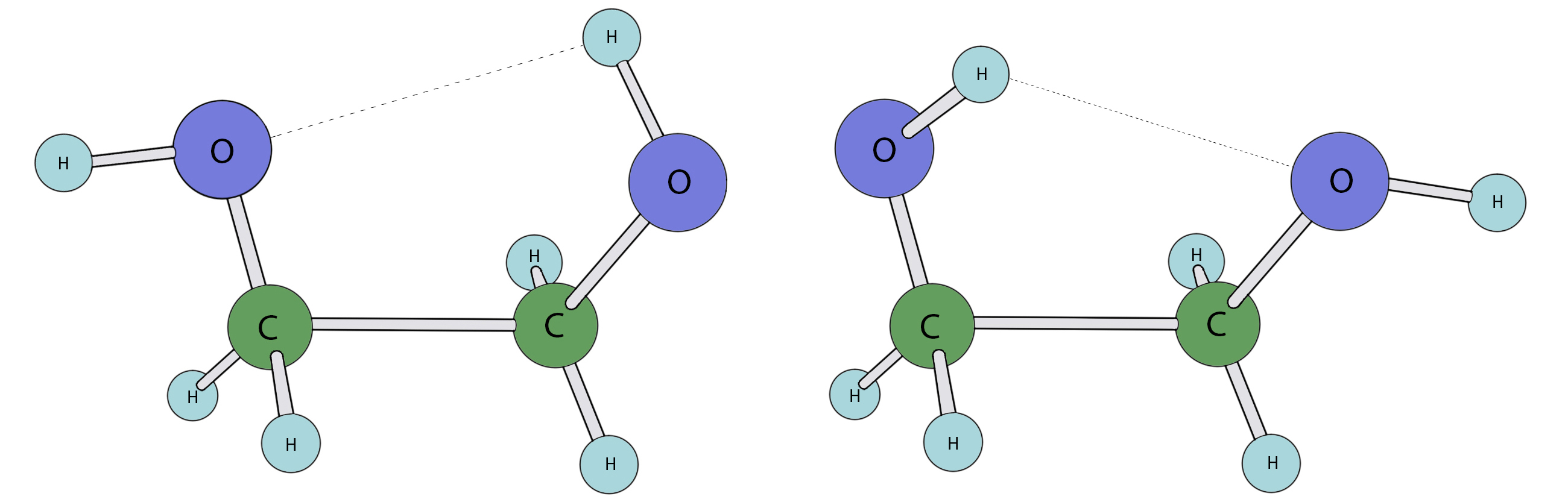}
\caption{Schematic molecules of ethylene glycol in two degenerate states of
the conformation $g'Ga$.} \label{fig:molecules}
\end{figure*}

Using spectra from extragalactic sources for
studying variation of constants was
first proposed by \citet{Sav56}. Later it was also shown \cite{Tho75} that
molecular spectra provide a way to determine possible variation of $\mu$. High
sensitivity for $\mu$ variation may exist in molecules which have more than
one equivalent potential minimum and which can tunnel between these minima
\cite{VKB04}. A well-known example of this kind of molecule is ammonia,
$\mathrm{NH_3}$, a compound fortunately abundant in interstellar medium (ISM).
Mixed tunneling-rotational transitions can be even more sensitive to the
change of $\mu$. But highly sensitive transitions of this type can be seen only in asymmetric isotopologues of
ammonia, $\mathrm{NH_2D}$ and $\mathrm{NHD_2}$ \cite{KLL10}.

Recently a large number of polyatomic molecules has been observed from the ISM
at the redshift $z=0.89$ \cite{MBG11}. That finding stimulated studies of the
molecules with mixed tunneling-rotational transitions. Up to now following
these molecules have been studied: hydronium ($\mathrm{H_3O^+}$) \cite{KPR11},
hydrogen peroxide ($\mathrm{H_2O_2}$) \cite{Koz11}, methanol ($\mathrm{CH_3OH}$) \cite{JXK11,LKR11}, methylamine ($\mathrm{CH_3NH_2}$)
\cite{IJKL12}, methyl mercaptan ($\mathrm{CH_3SH}$) \cite{JXKB13} and acetaldehyde ($\mathrm{C_2H_4O}$), acetamide ($\mathrm{CH_3CONH_2}$), methyl formate ($\mathrm{HCOOCH_3}$), acetic acid ($\mathrm{CH_3COOH}$) in \citet{JKX11}. The
strongest restriction for $\mu$ variation on a cosmological timescale
\eqref{lim_am} has been obtained by the observations of methanol spectra at
redshift $z = 0.89$ \cite{BJHB13}.

We suggest similar consideration of ethylene glycol $\mathrm{C_2 H_6 O_2}$. It
has two equivalent minima in the lowest $g'Ga$ conformation, see
\fref{fig:molecules}. Ethylene glycol has been detected in the comet C/1995 O1
(Hale-Bopp) \cite{CBBC04} and in ISM to the center of the Milky Way galaxy
\cite{HLJC02,RMMM08}. Also, recent success in detection of complex organic compounds
\cite{MBG11} at high redshift gives us hope to spot ethylene glycol outside of
the Milky Way. Therefore it is important to know which transitions in ethylene
glycol are especially sensitive to the change of $\mu$. In this paper we
calculate sensitivity coefficients for a large number of transitions,
including those which have not been observed yet.

\section{Method}

Let $\omega$ be a present-day experimentally observed transition frequency and
$\tilde{\omega}$ a frequency shifted due to possible time (and space) change
of $\alpha$ and $\mu$. This shift $\Delta\omega = \tilde{\omega} - \omega$ is
linked to $\Delta\alpha$ and $\Delta\mu$ through sensitivity coefficients
$Q_{\alpha}$ and $Q_{\mu}$ (we do not consider hyperfine transitions, which
may depend on additional parameters, such as nuclear $g$-factors):
\begin{equation}\label{meth1}
\frac{\Delta\omega}{\omega}
=Q_{\alpha}\frac{\Delta\alpha}{\alpha}+Q_{\mu}\frac{\Delta\mu}{\mu}\,.
\end{equation}
For tunneling-rotational spectra of molecules built of light elements ($Z< 10$) the sensitivity coefficient $Q_{\alpha}\ll 1$. At the same time typical coefficient $Q_{\mu}\gtrsim 1$. Therefore we neglect $\alpha$-dependence and link $\Delta\omega$ solely with $\Delta\mu$:
\begin{equation}\label{meth2}
\frac{\Delta\omega}{\omega}=Q_\mu\frac{\Delta\mu}{\mu}\,,
\end{equation}
\begin{equation}\label{meth3}
\Delta\mu=\tilde{\mu}-\mu\,,
\quad |\Delta\mu/\mu|\ll 1\,.
\end{equation}
Experimental data for the spectrum of ethylene glycol are taken from \citet{CCSL95}.

\subsection{Effective Hamiltonian}\label{SecHeff}

Ethylene glycol molecule has several conformations, which correspond to the
local minima of the potential. The lowest conformation is labeled as $g'Ga$
and is twofold degenerate \cite{CCSL95}. One can see from \fref{fig:molecules}
that two equivalent configurations differ mostly by the positions of the two
end OH groups. It is this conformation that has been observed in the ISM
\cite{HLJC02,CBBC04,RMMM08}. Below we discuss effective Hamiltonian for this
conformation.

Tunneling motion between two configurations of the $g'Ga$ conformation lifts
degeneracy and causes 7 GHz energy splitting of the ground state
\cite{HLJC02}. For a rotating molecule there is strong Coriolis interaction
between large amplitude tunneling mode and overall rotation \cite{CCSL95}. Our
main goal is to calculate sensitivity coefficients \eqref{meth2} for the
tunneling-rotational transitions. To this end we need to define how the
parameters of the effective Hamiltonian depend on the electron-to-proton mass
ratio $\mu$. This dependence can be reliably established only for the
relatively simple Hamiltonians \cite{LKR11,IJKL12}. More sophisticated
Hamiltonian can provide better accuracy for the transition frequencies, but do
not lead to significant improvement of the accuracy for the sensitivity
coefficients.

We found out that reasonable accuracy for the tunneling-rotational spectrum is
provided by the 14 parameter Hamiltonian, which in the molecular frame
$\xi,\eta,\zeta$ has the form:
\begin{subequations}\label{Heff1}
\begin{align}
  H_\mathrm{eff}
  &= C J_\xi^2 + B J_\eta^2 + A J_\zeta^2
  \label{Heff1a} \\
  &- \Delta_J \bm J^4 - \Delta_K J_\zeta^4 - \Delta_{JK} \bm J^2 J_\zeta^2
  \label{Heff1b} \\
  &+d_1 \bm J^2 \left(J_+^2 + J_-^2\right)
   +d_2 \left(J_+^4 + J_-^4\right)
  \label{Heff1c} \\
  &-\frac{\tau}{2}\left(F - W_C J_\xi^2 - W_B J_\eta^2 - W_A J_\zeta^2\right)
  \label{Heff1d} \\
  &+\left[d_3 J_\zeta + d_4 \left(J_+ + J_-\right)\right]\delta_{\tau',-\tau}
  \,.
  \label{Heff1e}
\end{align}
\end{subequations}
The first line corresponds to asymmetric top. For ethylene glycol $A\gg
B\gtrsim C$, so we can use a basis set $|J,K_A\rangle$ for prolate top with
$K_A \equiv \langle J_\zeta\rangle$. Lines (\ref{Heff1b},\ref{Heff1c})
describe diagonal and non-diagonal in $K_A$ centrifugal corrections
respectively. The line \eqref{Heff1d} describes tunneling degree of freedom,
where $F$ is tunneling frequency, $\tau=\pm1$ is tunneling quantum number.
Rotational constants weakly depend on the quantum number $\tau$ \cite{CCSL95},
so we define $A\equiv(A_{+1}+A_{-1})/2$ and $W_A\equiv(A_{+1}-A_{-1})/2$, etc.
Parameters $W_i$ can be considered as centrifugal corrections to the tunneling
frequency $F$ \cite{KPR11}. Finally, we introduced two terms \eqref{Heff1e},
which are non-diagonal in $\tau$ and depend on the rotational quantum numbers.
These terms describe Coriolis interaction between rotational degrees of
freedom and the tunneling mode. This interaction becomes particularly
important when levels with the same quantum number $J$, but different $\tau$
come close to each other. The term $d_3$ causes repulsion of such levels with
$\Delta K_A=0$ and the second one causes repulsion of levels with $\Delta
K_A=\pm1$. Addition of these two Coriolis terms to the Hamiltonian improves
quality of the fit by more than two orders of magnitude.\\[4mm]

\begin{longtable*}[c]{ccccrcccrrdd}
  \caption{Modelling of experimental spectrum of ethylene glycol from Ref.\
  \cite{CCSL95} with effective Hamiltonian \eqref{Heff1}. Frequencies are
  given in MHz and $\Delta\omega=\omega_\mathrm{theor}-\omega_\mathrm{exper}$.
  Quantum numbers $J,K_A,K_C$ correspond to the rigid asymmetric top and $v$
  is linked to $\tau$ from \eqref{Heff1d}: $v=(1-\tau)/2$. Unprimed and primed
  quantum numbers correspond to upper and lower states respectively.
  Sensitivity coefficients $\mathrm{Q}_\mu$ are calculated as described in
  \sref{SSecMuDep}. Estimated theoretical errors are given in parentheses. Two transitions observed to the Galactic center\cite{RMMM08} are in boldface.
}
  \label{tab1}
  \\
\hline\hline
 \multicolumn{1}{c}{$J$}
 &\multicolumn{1}{c}{$K_{A}$}
 &\multicolumn{1}{c}{$K_{C}$}
 &\multicolumn{1}{c}{$v\,$}
 &\multicolumn{1}{r}{$J'$}
 &\multicolumn{1}{c}{$K_{A}'$}
 &\multicolumn{1}{c}{$K_{C}'$}
 &\multicolumn{1}{c}{$v'$}
 &\multicolumn{1}{c}{$\omega_\mathrm{theor}$}
 &\multicolumn{1}{c}{$\omega_\mathrm{exper}$}
 &\multicolumn{1}{c}{$\Delta\omega$}
 &\multicolumn{1}{c}{$Q_{\mu}$}
 \\
 \hline
2   &   2   &   1   &   1   &   2   &   2   &   0   &   0   &   6889.3  &   6889.1  &   0.2 &   4.0(5)    \\
5   &   4   &   2   &   1   &   5   &   4   &   1   &   0   &   6952.6  &   6952.0  &   0.6 &   4.0(5)    \\
5   &   4   &   1   &   1   &   5   &   4   &   2   &   0   &   6954.6  &   6953.3  &   1.3 &   4.0(5)    \\
3   &   3   &   1   &   1   &   3   &   3   &   0   &   0   &   6956.9  &   6957.2  &   -0.3    &   4.0(5)    \\
3   &   3   &   0   &   1   &   3   &   3   &   1   &   0   &   6962.9  &   6963.8  &   -0.9    &   4.0(5)    \\
4   &   4   &   1   &   1   &   4   &   4   &   0   &   0   &   6964.1  &   6963.9  &   0.2 &   4.0(5)    \\
4   &   4   &   0   &   1   &   4   &   4   &   1   &   0   &   6964.3  &   6963.9  &   0.4 &   4.0(5)    \\
4   &   3   &   1   &   1   &   4   &   3   &   2   &   0   &   6972.4  &   6972.4  &   0.0 &   4.0(5)    \\
5   &   3   &   2   &   1   &   5   &   3   &   3   &   0   &   7024.6  &   7024.7  &   -0.1    &   4.0(5)    \\
2   &   2   &   0   &   1   &   2   &   2   &   1   &   0   &   7026.3  &   7026.5  &   -0.2    &   4.0(5)    \\
5   &   1   &   4   &   0   &   5   &   1   &   5   &   1   &   7600.7  &   7600.7  &   0.0 &   -1.7(4)   \\
1   &   1   &   0   &   1 &\quad 1  &   1   &   1   &  0 &\quad  7925.6 &   7925.5  &   0.1 &   3.6(4)    \\
4   &   2   &   2   &   1   &   4   &   2   &   3   &   0   &   7949.0  &   7948.9  &   0.1 &   3.6(4)    \\
5   &   2   &   3   &   1   &   5   &   1   &   4   &   0   &   9217.5  &   9217.4  &   0.1 &   3.3(4)    \\
2   &   1   &   1   &   1   &   2   &   1   &   2   &   0   &   9852.3  &   9852.1  &   0.2 &   3.1(3)    \\
2   &   0   &   2   &   1   &   1   &   1   &   1   &   1   &   10534.7 &   10534.5 &   0.2 &   0.97(1)    \\
2   &   0   &   2   &   0   &   1   &   1   &   1   &   0   &   10551.8 &   10551.9 &   -0.1    &   1.00(1)    \\
1   &   1   &   0   &   0   &   1   &   0   &   1   &   0   &   10747.6 &   10747.5 &   0.1 &   1.00(1)    \\
1   &   1   &   0   &   1   &   1   &   0   &   1   &   1   &   10754.2 &   10754.3 &   -0.1    &   1.01(1)    \\
5   &   1   &   5   &   1   &   4   &   2   &   2   &   1   &   11488.2 &   11488.0 &   0.2 &   0.99(1)    \\
5   &   1   &   5   &   1   &   5   &   0   &   5   &   0   &   11745.2 &   11745.0 &   0.2 &   2.8(3)    \\
2   &   1   &   1   &   0   &   2   &   0   &   2   &   0   &   11785.9 &   11785.8 &   0.1 &   1.00(1)    \\
2   &   1   &   1   &   1   &   2   &   0   &   2   &   1   &   11810.2 &   11810.3 &   -0.1    &   1.03(1)    \\
2   &   1   &   2   &   0   &   1   &   1   &   1   &   1   &   12492.6 &   12492.7 &   -0.1    &   2.4(3)    \\
3   &   1   &   2   &   1   &   3   &   1   &   3   &   0   &   12689.2 &   12689.1 &   0.1 &   2.4(2)    \\
$\mathbf{2}$   &   $\mathbf{0}$   &   $\mathbf{2}$   &   $\mathbf{0}$   &   $\mathbf{1}$   &   $\mathbf{0}$   &   $\mathbf{1}$   &   $\mathbf{1}$   &   $\mathbf{13380.4}$ &   $\mathbf{13380.6}$ &   \bf -0.\bf 2    &  \bf -0.\bf 6(2)   \\
3   &   1   &   2   &   0   &   3   &   0   &   3   &   0   &   13444.8 &   13444.8 &   0.0 &   1.03(1)    \\
3   &   1   &   2   &   1   &   3   &   0   &   3   &   1   &   13571.4 &   13571.5 &   -0.1    &   1.2(2)    \\
1   &   1   &   0   &   0   &   0   &   0   &   0   &   1   &   13990.8 &   13990.9 &   -0.1    &   -0.5(2)   \\
2   &   1   &   1   &   0   &   1   &   1   &   0   &   1   &   14412.1 &   14412.2 &   -0.1    &   -0.4(2)   \\
4   &   1   &   3   &   1   &   3   &   2   &   2   &   1   &   14678.1 &   14678.1 &   0.0 &   1.00(1)    \\
4   &   1   &   3   &   0   &   3   &   2   &   2   &   0   &   14706.0 &   14706.2 &   -0.2    &   1.02(1)    \\
4   &   1   &   3   &   1   &   4   &   0   &   4   &   1   &   15808.4 &   15809.0 &   -0.6    &   1.1(1)    \\
4   &   1   &   3   &   0   &   4   &   0   &   4   &   0   &   15972.1 &   15971.9 &   0.2 &   1.01(1)    \\
1   &   1   &   1   &   1   &   1   &   0   &   1   &   0   &   16734.2 &   16734.1 &   0.1 &   2.2(2)    \\
4   &   1   &   3   &   1   &   4   &   1   &   4   &   0   &   16786.7 &   16786.8 &   -0.1    &   2.1(2)    \\
5   &   2   &   4   &   0   &   5   &   1   &   4   &   1   &   16897.4 &   16897.6 &   -0.2    &   -0.2(2)   \\
1   &   0   &   1   &   1   &   0   &   0   &   0   &   0   &   17153.8 &   17153.6 &   0.2 &   2.2(2)    \\
1   &   1   &   1   &   1   &   0   &   0   &   0   &   1   &   19977.4 &   19977.5 &   -0.1    &   1.00(1)    \\
1   &   1   &   1   &   0   &   0   &   0   &   0   &   0   &   19982.4 &   19982.3 &   0.1 &   1.00(1)    \\
$\mathbf{3}$   &   $\mathbf{0}$   &   $\mathbf{3}$   &   $\mathbf{0}$   &  $\mathbf{2}$   &   $\mathbf{0}$   &   $\mathbf{2}$   &   $\mathbf{1}$   &   $\mathbf{23392.9}$ &   $\mathbf{23393.0}$ &  \bf -0.\bf 1   &  \bf 0.\bf 1(1)    \\
3   &   1   &   2   &   0   &   2   &   1   &   1   &   1   &   25027.5 &   25027.6 &   -0.1    &   0.2(1)    \\
3   &   2   &   1   &   1   &   3   &   1   &   2   &   1   &   28259.1 &   28259.3 &   -0.2    &   1.00(1)    \\
3   &   2   &   1   &   0   &   3   &   1   &   2   &   0   &   28292.2 &   28291.9 &   0.3 &   0.98(1)    \\
4   &   0   &   4   &   0   &   3   &   0   &   3   &   1   &   33272.8 &   33272.9 &   -0.1    &   0.5(1)    \\
3   &   2   &   2   &   0   &   3   &   1   &   3   &   0   &   33656.7 &   33656.4 &   0.3 &   0.9(1)    \\
4   &   2   &   3   &   0   &   3   &   2   &   2   &   1   &   33806.6 &   33806.3 &   0.3 &   0.4(1)    \\
4   &   3   &   2   &   0   &   3   &   3   &   1   &   1   &   33977.3 &   33976.8 &   0.5 &   0.4(1)    \\
4   &   3   &   1   &   0   &   3   &   3   &   0   &   1   &   33995.2 &   33994.6 &   0.6 &   0.4(1)    \\
4   &   1   &   3   &   0   &   3   &   1   &   2   &   1   &   35673.5 &   35673.4 &   0.1 &   0.4(1)    \\
4   &   2   &   3   &   0   &   4   &   1   &   4   &   0   &   35915.2 &   35915.1 &   0.1 &   0.9(1)    \\
3   &   1   &   3   &   1   &   2   &   1   &   2   &   0   &   36061.4 &   36061.5 &   -0.1    &   1.6(1)    \\
3   &   0   &   3   &   1   &   2   &   0   &   2   &   0   &   37188.1 &   37187.7 &   0.4 &   1.5(1)   \\
3   &   2   &   2   &   1   &   2   &   2   &   1   &   0   &   37557.0 &   37557.2 &   -0.2    &   1.6(1)    \\
3   &   1   &   3   &   1   &   2   &   0   &   2   &   1   &   38019.3 &   38019.7 &   -0.4    &   1.00(1)    \\
3   &   1   &   3   &   0   &   2   &   0   &   2   &   0   &   38070.3 &   38070.2 &   0.1 &   1.1(1)    \\
5   &   2   &   4   &   1   &   5   &   1   &   5   &   1   &   38351.8 &   38351.8 &   0.0 &   1.00(1)    \\
3   &   1   &   2   &   1   &   2   &   1   &   1   &   0   &   38973.6 &   38973.4 &   0.2 &   1.5(1)    \\
5   &   0   &   5   &   0   &   4   &   0   &   4   &   1   &   42628.9 &   42629.1 &   -0.2    &   0.6(1)    \\
5   &   2   &   4   &   0   &   4   &   2   &   3   &   1   &   43919.9 &   43919.7 &   0.2 &   0.5(1)    \\
5   &   4   &   2   &   0   &   4   &   4   &   1   &   1   &   44199.5 &   44199.7 &   -0.2    &   0.5(1)    \\
5   &   4   &   1   &   0   &   4   &   4   &   0   &   1   &   44200.4 &   44200.6 &   -0.2    &   0.5(1)    \\
5   &   3   &   3   &   0   &   4   &   3   &   2   &   1   &   44264.4 &   44263.9 &   0.5 &   0.5(1)    \\
5   &   3   &   2   &   0   &   4   &   3   &   1   &   1   &   44326.3 &   44326.0 &   0.3 &   0.5(1)    \\
5   &   2   &   3   &   0   &   4   &   2   &   2   &   1   &   45180.0 &   45179.7 &   0.3 &   0.5(1)    \\
4   &   1   &   4   &   1   &   3   &   1   &   3   &   0   &   45547.5 &   45547.6 &   -0.1    &   1.4(1)    \\
5   &   1   &   4   &   0   &   4   &   1   &   3   &   1   &   46166.4 &   46166.5 &   -0.1    &   0.6(1)    \\
4   &   0   &   4   &   1   &   3   &   0   &   3   &   0   &   47217.7 &   47217.4 &   0.3 &   1.4(1)    \\
4   &   2   &   3   &   1   &   3   &   2   &   2   &   0   &   47696.1 &   47696.4 &   -0.3    &   1.4(1)    \\
4   &   3   &   2   &   1   &   3   &   3   &   1   &   0   &   47888.3 &   47888.7 &   -0.4    &   1.4(1)    \\
4   &   3   &   1   &   1   &   3   &   3   &   0   &   0   &   47906.6 &   47906.9 &   -0.3    &   1.4(1)    \\
4   &   2   &   2   &   1   &   3   &   2   &   1   &   0   &   48366.6 &   48366.8 &   -0.2    &   1.4(1)    \\
4   &   1   &   3   &   0   &   3   &   0   &   3   &   1   &   49244.9 &   49245.2 &   -0.3    &   0.6(1)    \\
4   &   1   &   3   &   1   &   3   &   1   &   2   &   0   &   49581.3 &   49581.1 &   0.2 &   1.4(1)    \\
5   &   3   &   2   &   1   &   5   &   2   &   3   &   1   &   49690.5 &   49690.6 &   -0.1    &   1.00(1)    \\
3   &   1   &   2   &   1   &   2   &   0   &   2   &   0   &   50759.5 &   50759.3 &   0.2 &   1.4(1)    \\
5   &   1   &   5   &   1   &   4   &   1   &   4   &   0   &   55352.4 &   55352.7 &   -0.3    &   1.3(1)    \\
5   &   2   &   4   &   1   &   4   &   2   &   3   &   0   &   57789.0 &   57789.2 &   -0.2    &   1.4(1)    \\
5   &   3   &   2   &   1   &   4   &   3   &   1   &   0   &   58219.2 &   58219.4 &   -0.2    &   1.4(1)    \\
5   &   2   &   3   &   1   &   4   &   2   &   2   &   0   &   59076.2 &   59076.3 &   -0.1    &   1.4(1)    \\

\hline\hline
\end{longtable*}

\subsection{Determining $\mu$-dependence of the effective
Hamiltonian}\label{SSecMuDep}

Below we find parameters of the effective Hamiltonian \eqref{Heff1} from the
fit to the experimental spectrum \cite{CCSL95}. However, we assume that, in
principle, these parameters can be found from the \textit{ab initio}
calculations. Repeating such calculations with different values of $\mu$ we
can find $\mu$-dependence of our parameters. On the other hand, at least for
the largest parameters of our Hamiltonian we can find approximate
$\mu$-dependence without extensive calculations. For example, rotational
constants scale as $1/(MR^2)$, where $M$ is nuclear mass and $R$ is
equilibrium internuclear distance. This means that in atomic units, which are
traditionally used to define sensitivity coefficients, the rotational
constants $A,\,B,\,C$ scale as $\mu^1$. Similar arguments show that
centrifugal corrections $\Delta_J,\,\Delta_K,\,\Delta_{JK}$ and $d_1,\,d_2$
scale as $\mu^2$. The accuracy of these scalings is on the order of 1\%, or so
\cite{LKR11,IJKL12}. Without calculating these scalings more accurately we can
not improve the accuracy for the sensitivity coefficients $Q_\mu$ by adding
higher centrifugal corrections to our Hamiltonian.

In order to find $\mu$-dependence of the constant $F$ we can either do some
model calculations for the tunneling mode \cite{FK07a,KPR11}, or use
experimental data for the deuterated species \cite{VKB04}. We use the latter
approach here. Using semiclassical arguments we can expect following scaling
of the parameter $F$:
\begin{align}\label{Fscal1}
 &F = a\mu^{1/2} \exp\left(-b/\mu^{1/2}\right)\,,
 \\
 \label{Fscal2}
 &Q_\mu(F)=\frac12 \left(1+b/\mu^{1/2}\right)\,.
\end{align}
We can find parameters $a$ and $b$ from experimental values of $F$ for two
isotopologues of ethylene glycol: $F=6958$~MHz for HOCH$_2$CH$_2$OH and
$F=293$~MHz for DOCH$_2$CH$_2$OD \cite{CCSL95}. We consider deuterated
molecule as one with a proton of a double mass.

According to \fref{fig:molecules} two degenerate configurations differ mostly
by the positions of the OH (or OD) groups. We do not know the exact tunneling path and
the respective effective tunneling mass. In the two limiting models the
tunneling motion can be approximated as a rotation of the rigid OH groups, or
simply as the motion of two hydrogen atoms. In the first case the tunneling
masses for two isotopologues are $M_1=M_{\rm H}M_{\rm O}/(M_{\rm H}+M_{\rm
O})$ and $M_2=M_{\rm D}M_{\rm O}/(M_{\rm D}+M_{\rm O})$. In the second case
$M_1=M_{\rm H}$ and $M_2=M_{\rm D}$. For the first case we get $Q_\mu(F)=4.31$
and for the second case $Q_\mu(F)=3.91$. Actual tunneling mass should lie
between these two limiting cases, so the conservative estimate is:
\begin{align}\label{Fscal3}
 Q_\mu(F)= 4.1\pm0.2\,.
\end{align}

Finally we need to determine $\mu$-dependence for centrifugal corrections to
the tunneling frequency $W_i$ and Coriolis parameters $d_3$ and $d_4$. At
present there is no accurate theory for these terms, but it is usually assumed
\cite{JXK11,IJKL12} that their scaling with $\mu$ is given by a
product of lower-order Hamiltonian terms, in this case the tunneling
and rotational constants. The sensitivity coefficients for the
higher-order terms are thus given by
\begin{align}\label{Fscal4}
 Q_\mu(W_i)= Q_\mu(d_3) = Q_\mu(d_4) = Q_\mu(F)+1\,.
\end{align}

Knowing the scalings of the parameters of the effective Hamiltonian we can
find $\mu$ dependence of the transition frequencies by diagonalizing
$H_\mathrm{eff}$ for several sets of parameters, which corresponds to
different values of $\mu$.

\begin{table}[htb]
\caption{Optimized parameters of the effective Hamiltonian \eqref{Heff1} in
MHz.} \label{tab3}
\begin{tabular}[c]{ld}
\hline\hline
$A$                & 15363.284    \\    
$B$                & 5587.121     \\    
$C$                & 4613.531     \\    
$\Delta_J$         & 0.0074       \\    
$\Delta_K$         & 0.0774       \\    
$\Delta_{JK}\quad$ & -0.0329      \\    
$d_1$              & 0.0025       \\    
$d_2$              & 0.00003      \\    
$F$                & 6958.1       \\    
$W_C$              & 13.088       \\    
$W_B$              & -0.425       \\    
$W_A$              & 2.771        \\    
$d_3$              & 217.09       \\    
$d_4$              & 50.83        \\    
\hline\hline
\end{tabular}
\end{table}

\section{Numerical results and discussion}\label{SecResults}
\begin{figure*}[tbh]
\includegraphics[width=0.8\textwidth]{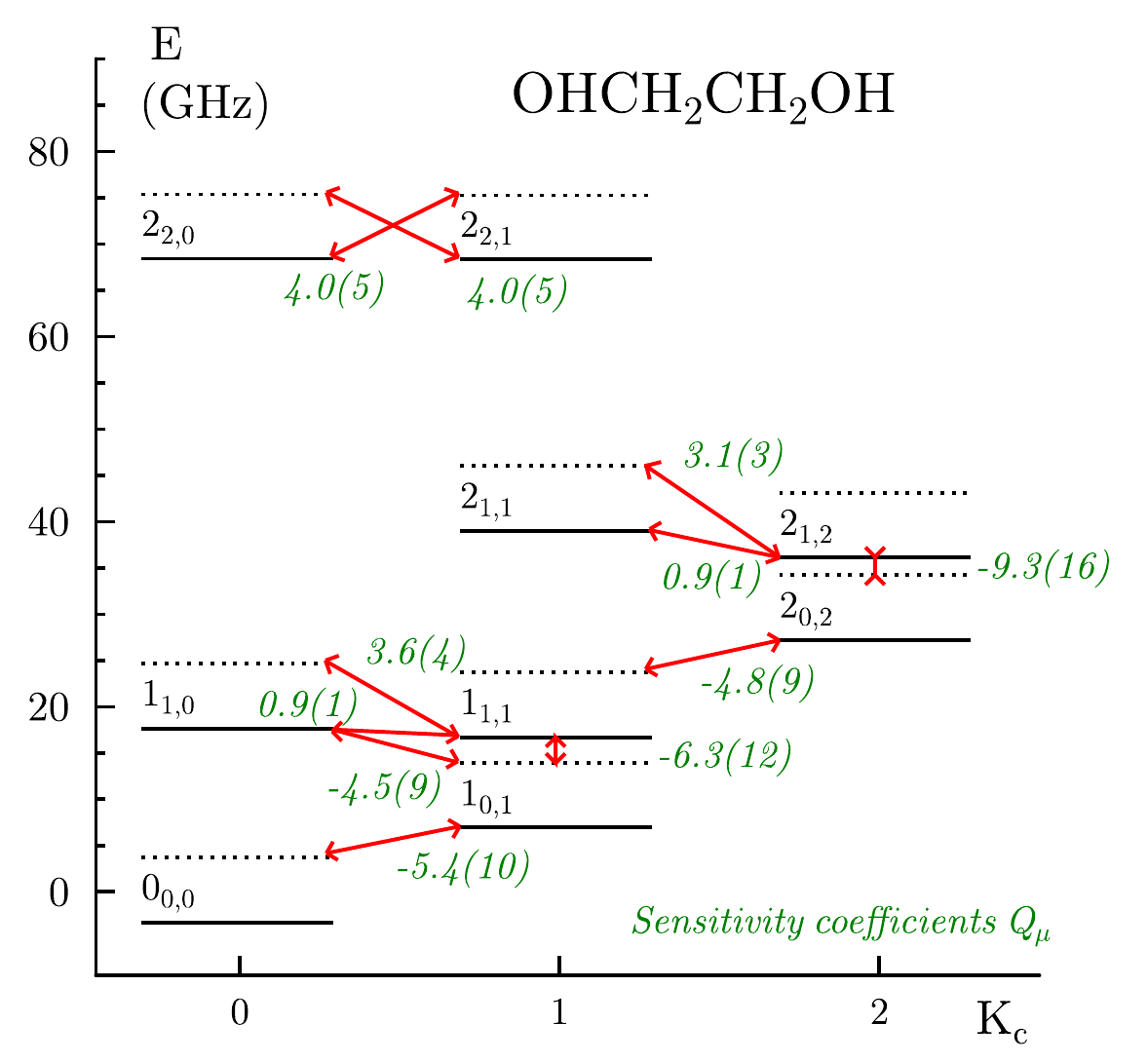}
\caption{The lowest part of the tunneling-rotational spectrum of ethylene glycol. Quantum numbers of energy levels are given as $J_{K_a,K_c}$.
Solid lines correspond to $v=0$, dotted ones to $v=1$.  Sensitivity
coefficients $Q_\mu$ are in italic. Note that predominantly rotational transitions
($\Delta v=0$) have $Q_\mu \approx 1$.} \label{fig:energies}
\end{figure*}

The lowest part of the tunneling-rotational spectrum of ethylene glycol is
shown in \fref{fig:energies}. Effective Hamiltonian discussed above is used to
model the spectrum and to calculate corresponding sensitivities $Q_\mu$. We
fit 14 parameters from \Eref{Heff1} to the experimental transition frequencies
measured by \citet{CCSL95}. Typical temperature of the ISM is $T\lesssim 10$K,
so the levels with $J>5$ are weakly populated. Transitions for higher $J$'s
observed in \cite{HLJC02} correspond to the much warmer gas and are very broad
(linewidths $>$ 20 km/s). Because of that we use Hamiltonian, which includes
only lowest centrifugal corrections and restrict our consideration to the
levels with $J\le 5$. Results of this fit are presented in \tref{tab1}.
Achieved agreement is quite satisfactory taking into account relative
simplicity of the model we use. The maximum deviation $\Delta\omega$ from the
measured frequency is 1.3 MHz, while the \textit{rms} deviation is about 0.3
MHz.

The optimized parameters of the model are listed in \tref{tab3}. One can see
that centrifugal and Coriolis corrections to the tunneling are rather large.
The largest term $d_3$ causes up to a hundred MHz shifts of several transition
lines. Being nondiagonal in the quantum number $\tau$ the terms $d_3$ and
$d_4$ become important only for the close levels with the opposite values of
$\tau$ and the same $J$. There are only several such levels with $J\le 5$.
Other terms of the Hamiltonian contribute more uniformly to the
tunneling-rotational spectrum of the molecule.

\begin{table}[htb]
\caption{Predicted low-frequency tunneling-rotational transitions (in the
range from 800 MHz to 6.8 GHz) and their sensitivity coefficients $Q_\mu$.
Frequencies are in MHz and wavelengths are in cm. Other notations are the same
as in \tref{tab1}.} \label{tab2}
\begin{tabular}[c]{ccccrcccrrd}
\hline\hline
 \multicolumn{1}{c}{$J$}
 &\multicolumn{1}{c}{$K_{A}$}
 &\multicolumn{1}{c}{$K_{C}$}
 &\multicolumn{1}{c}{$v\,$}
 &\multicolumn{1}{r}{$J'$}
 &\multicolumn{1}{c}{$K_{A}'$}
 &\multicolumn{1}{c}{$K_{C}'$}
 &\multicolumn{1}{c}{$v'$}
 &\multicolumn{1}{c}{$\omega$}
 &\multicolumn{1}{c}{$\lambda$}
 &\multicolumn{1}{c}{$Q_{\mu}$}
 \\
\hline
3   &   1   &   3   &   0   &   3   &   0   &   3   &   1   &    882.2  &     34.0   &   -16.5(58)  \\
1   &   1   &   0   &   0 &\quad 1  &   1   &   1   &   0   &    966.4  &    ~31.0   &   0.9(1)    \\
4   &   0   &   4   &   1   &   4   &   1   &   4   &   0   &    978.3  &     30.6   &   17.8(38)   \\
4   &   2   &   2   &   0   &   4   &   2   &   3   &   0   &   1000.9  &     30.0   &   0.95(1)    \\
3	&	1	&	3	&	1	&	3	&	1	&	2	&	0	&	1181.6	&	25.4	&	17.9(26)	\\
2   &   1   &   2   &   0   &   2   &   0   &   2   &   1   &   1957.9  &     15.3   &   -9.3(16)   \\
3   &   3   &   1   &   0   &   4   &   2   &   2   &   1   &   2641.3  &      11.4   &   -6.9(12)   \\
3   &   1   &   2   &   0   &   2   &   2   &   1   &   0   &   2653.8  &      11.3   &   1.2(2)   \\
3   &   1   &   2   &   1   &   2   &   2   &   1   &   1   &   2682.2  &      11.2   &   0.98(2)   \\
4   &   1   &   3   &   0   &   4   &   1   &   4   &   1   &  ~2815.2  &     10.7   &   -6.2(11)   \\
1   &   1   &   1   &   0   &   1   &   0   &   1   &   1   &   2828.6  &     10.6   &   -6.3(12)   \\
2   &   1   &   1   &   0   &   2   &   1   &   2   &   0   &   2892.1  &     10.4   &   0.9(1)    \\
1   &   0   &   1   &   0   &   0   &   0   &   0   &   1   &   3243.2  &      9.2   &   -5.4(10)   \\
2   &   0   &   2   &   0   &   1   &   1   &   1   &   1   &   3598.8  &      8.3   &   -4.8(9)   \\
1   &   1   &   0   &   0   &   1   &   0   &   1   &   1   &   3795.0  &      7.9   &   -4.5(9)   \\
2	&	1	&	2	&	1	&	2	&	1	&	1	&	0	&	4043.9	&	7.4	&	6.2(8)	\\
2   &   2   &   1   &   1   &   3   &   1   &   2   &   0   &   4303.7  &      7.0   &   5.8(7)   \\
2	&	1	&	1	&	0	&	2	&	0	&	2	&	1	&	4850.0	&	6.2	&	-3.2(7)	\\
3	&	1	&	2	&	0	&	3	&	1	&	3	&	0	&	5703.3	&	5.3	&	0.6(5)	\\
4	&	2	&	3	&	1	&	4	&	2	&	2	&	0	&	5938.8	&	5.0	&	4.5(5)	\\
1	&	1	&	1	&	1	&	1	&	1	&	0	&	0	&	5986.6	&	5.0	&	4.5(5)	\\
4	&	1	&	4	&	0	&	4	&	0	&	4	&	0	&	6107.3	&	4.9	&	1.4(3)	\\
3	&	1	&	2	&	0	&	3	&	0	&	3	&	1	&	6585.5	&	4.6	&	-1.7(4)	\\
3	&	2	&	2	&	1	&	3	&	2	&	1	&	0	&	6611.0	&	4.5	&	4.2(5)	\\

\hline

\hline\hline
\end{tabular}
\end{table}

In the experiment \cite{CCSL95} only transitions above 6.8 GHz were detected.
At the same time, we are primarily interested in low frequency mixed
tunneling-rotational transitions with $\omega \gtrsim 1$~GHz, where high
sensitivities are possible. Transitions with even lower frequencies are hardly
detectable by modern Earth based radio telescopes. To the best of our
knowledge such transitions for ethylene glycol were never seen. Thus, we used
our effective Hamiltonian with optimal parameters from \tref{tab3} to search
for such low frequency transitions. We again restricted ourselves to $J\le 5$
where our model has been tested against the experiment and proved to be quite
reliable. Within these limits we found 24 low frequency
transitions listed in \tref{tab2}. Some of them are shown in
\fref{fig:energies}. We estimate the accuracy of the predicted frequencies to
be about 0.5 MHz.

We used the same optimal effective Hamiltonian to calculated sensitivities
$Q_\mu$ for all transitions from Tables \ref{tab1} and \ref{tab2} as was
described in \sref{SSecMuDep}. In order to estimate the uncertainty for the
obtained $Q$ factors we made several additional calculations. The main error
comes from the uncertainty \eqref{Fscal3}, so we first calculated
sensitivities for maximum and minimum values of $Q_\mu(F)$.

As we pointed out above, the theoretical grounds for \Eref{Fscal4} are not
very solid. Therefore we repeated calculations of the $Q$ factors with smaller
number of fitting parameters. In particular, we successively turned each of
the parameters $d_i$ and $W_i$ to zero and made fits for remaining 13
parameter sets. After that we calculated sensitivity coefficients $Q_\mu$ for
such restricted parameter sets. Note that according to \Eref{meth2} the value
of $Q_\mu$ is inversely proportional to the transition frequency $\omega$. For
the low frequency transitions predicted frequency may be quite sensitive to
the values of the parameters and can be significantly different for the best
fit and for the restricted fits. This part of the error is trivial and can be
easily eliminated for example by using the experimental frequencies instead of
the calculated ones. Because of that we excluded this error by using
frequencies from the best fit in the denominator, while the frequency shift
$\Delta\omega$ in the numerator was recalculated for each restricted set of
parameters.

After all these calculations being done we estimate the error $\Delta Q_\mu$
for each transition by taking maximum deviation from the main calculation with
optimal parameters. In most cases maximum error comes from the uncertainty in
the value of $Q_\mu(F)$. However, for some important transitions with high
sensitivities the largest deviation corresponds to the fit with $d_4$ set to
zero. The optimal value of this parameter is rather large and setting it to
zero significantly influences both the frequencies and their $\mu$ dependence.
The error associated with the parameter $d_4$ is particularly large for some
of the most interesting low-frequency transitions. On the contrary,
transitions with sensitivities close to unity are not sensitive to any changes
of the parameters discussed above. Here the main error comes from the
uncertainty to which we know $\mu$ dependence of the rotational constants. In
Refs.\ \cite{LKR11,IJKL12} this error was estimated to be about 1\%. This is
the minimal error of our calculation for the predominantly rotational
transitions with $Q_\mu \approx 1$.

\subsection*{Conclusion}

During last few years the molecules with mixed tunneling-rotational spectra
proved to be very useful for constraining possible $\mu$ variation on the
large space-time scale. Current most stringent limit on such variation has
been obtained with methanol \cite{BJHB13}. There is large variability in the
abundances of different species in the ISM and in observed intensities of
different molecular lines. Because of that it is useful to study all
potentially interesting molecules and transitions. In this paper we considered
one of the last unstudied relatively simple molecules with the tunneling mode.

We found several low frequency transitions in the range between 0.8 and 7 GHz
with high sensitivity to $\mu$ variation of both signs. Note that it is the
difference in sensitivities that is important for the observation of
$\mu$-variation. The maximum difference in sensitivities for these transitions
is close to 34. This is comparable to the differences earlier found for
methanol \cite{JXK11,LKR11}. For a higher frequencies there are several
transitions around 7.0 GHz with sensitivities $Q_\mu\approx 4.0$ and one
transition at 7.6 GHz with sensitivity $Q_\mu\approx -1.7$. Small frequency
differences may help to observe these lines simultaneously, minimizing
possible systematic errors.

Ethylene glycol has been detected in the ISM \cite{HLJC02,RMMM08}, which makes
it one of the perspective candidates for the search for $\mu$ variation.
Observed lines from the cold molecular clouds in the Milky Way can be very
narrow allowing for high precision spectroscopy. This can be used
\cite{MLK09,LKR11,EVB11} to study the possible dependence of the
electron-to-proton mass ratio on the local matter density, which is predicted
by models with chameleon scalar fields \cite{KW04,OP08}. At the same time high
redshift observations of the tunneling-rotational lines can be used to study
$\mu$ variation on the cosmological timescale \cite{EVBL12,BJHB13}.\\[2mm]


We thank Sergei Levshakov for helpful discussions. This work is partly
supported by the Russian Foundation for Basic Research, Grant No.\
14-02-00241.

%

\end{document}